# Multicenter automatic detection of invasive carcinoma on breast whole slide images

Rémy Peyret PhD[1], Nicolas Pozin PhD[1], Stéphane Sockeel PhD[1], Solène-Florence Kammerer-Jacquet MD[2], Julien Adam MD-PhD[3], Claire Bocciarelli MD[4], Yoan Ditchi MD[5], Christophe Bontoux MD[6], Thomas Depoilly MD[7], Loris Guichard MD[8], Elisabeth Lanteri MD[9], Marie Sockeel MD[1,10], Sophie Prévot MD-PhD[8]

[1] Primaa, Paris, France

[2] Pathologist, Rennes University hospital, France

[3] Pathologist, Paris Saint-Joseph hospital, Paris, France

[4] Pathologist, Brest university hospital, France

[5] Pathologist, Saint-Antoine hospital, APHP, Paris, France.

[6] Pathologist, Henri-Mondor hospital, APHP, Paris, France

[7] Pathologist, René Dubos hospital, Pontoise, France

[8] Pathologist, Bicêtre hospital, APHP, Paris Saclay University, France

[9] Pathologist, Nice University hospital, France

[10] Pathologist, Primaa, Paris, France

**Corresponding author:**

Name, Address : Rémy Peyret, PhD[1]; 23 rue des Jeuneurs, Paris, France

Email: correspondence@primaalab.com

ABSTRACT

**Background**

Breast cancer is one of the most prevalent cancers worldwide and pathologists are closely involved in establishing a diagnosis. Tools to assist in making a diagnosis are required to manage the increasing workload. In this context, artificial intelligence (AI) and deep-learning based tools may be used in daily pathology practice. However, it is challenging to develop fast and reliable algorithms that can be trusted by practitioners, whatever the medical center.

**Methods**

We describe a patch-based algorithm that incorporates a convolutional neural network to detect and locate invasive carcinoma on breast whole-slide images. The network was trained on a dataset extracted from a reference acquisition center. We then performed a calibration step based on transfer learning to maintain the performance when translating on a new target acquisition center by using a limited amount of additional training data. Performance was evaluated using classical binary measures (accuracy, recall, precision) for both centers (referred to as "test reference dataset" and "test target dataset") and at two levels: patch and slide level.

**Findings**

At patch level, accuracy, recall, and precision of the model on the reference and target test sets were 92.1% and 96.3%, 95% and 87.8%, and 73.9% and 70.6%, respectively. At slide level, accuracy, recall, and precision were 97.6% and 92.0%, 90.9% and 100%, and 100% and 70.8% for test sets 1 and 2, respectively.

**Interpretation**

The high performance of the algorithm at both centers shows that the calibration process is efficient. This is performed using limited training data from the new target acquisition center and requires that the model is trained beforehand on a large database from a reference center. This methodology allows the implementation of AI diagnostic tools to help in routine pathology practice.

**Financial Disclosure**

This study was carried out at Primaa, which is a startup company. The funders had no role in study design, data collection and analysis, decision to publish, or preparation of the manuscript. Two of the authors, M.S. and S.S., are co-founders and employees of this company, two authors, R.P. and N.P., are employees, and the rest are collaborators but did not receive any salary from Primaa.

**INTRODUCTION**

Cancer detection is a major public health issue, with almost 10 million cancer deaths worldwide in 2020, 19.3 million new cases diagnosed, and an expected rise to 28.4 million cases from 2020 to 2040 (+47%).[1] Breast cancer (BC) has now surpassed lung cancer as the most commonly diagnosed malignancy (2.3 million new cases diagnosed worldwide, 11.7% of all cancer diagnoses) and is the leading or second cause of premature death in women in many countries according to the World Health Organization. Accurate and prompt detection of BC is essential to improve treatment efficacy and survival.

Current diagnosis of BC relies on close visual examination of surgical or biopsy material at cellular level by highly qualified pathologists. The average annual workload of pathologists has increased by around 5–10%[2] and current data indicate a shortage of histopathologists worldwide[3] leading to overwork, fatigue, and a higher risk of mistakes and diagnostic errors.[2] Furthermore, the recent Covid-19 pandemic has led to a significant delay and backlog in cancer diagnosis with the number of new cases diagnosed falling by 23.3% in 2020 according to the National Cancer Institute.[4]

An increasing number of histopathology departments are going digital to overcome this problem.[5] Recent advances in slide scanners and information technology infrastructure enable the use of high-resolution digitized images called whole slide images (WSI) in pathologists daily routine. Digital pathology provides telemedicine, slide sharing for collaboration or second opinions, easier generation of reports, and long-term slide preservation. Above all, digital pathology allows the development of computational pathology (i.e., the analysis of WSI by algorithms in order to help pathologists in their diagnosis). This new field resulting from computer vision has recently come to the attention of the pathology community by promising time saving, better reproducibility, and better accuracy.[6] The use of advanced machine learning and deep-learning has been demonstrated on breast,[7,8] prostate,[9,10] skin,[11] lung,[12] and colorectal[13] cancer detection in numerous studies.

Several studies[14-17] have shown that machine learning based classifiers are built and used directly on sub-images extracted from WSI to predict whether they contain cancer or not. However, in a clinical routine context, a pipeline locating cancer on raw WSI is needed.

Few studies take into account computing times. In the study carried out by Pantanowitz et al.[18] a complete pipeline was proposed on prostate tissues, but computation times were approximately 20 min per slide and were therefore not compatible with daily clinical practice. In this study, we propose a fast end-to-end pipeline from raw WSI to cancer location meeting the needs of routine pathology practice.

Another requirement when designing a tool for clinical practice is the ability to maintain performance when transferring to a new medical center. Slide preparation and acquisition scanners might differ, leading to various image aspects. An algorithm that has learnt on data from a given acquisition center can be very efficient on that specific acquisition center, while performance may deteriorate when moving to a new center. In a previous study,[19] the authors used a deep-learning model for style normalization before their classification model in order to make their system center agnostic. In another study,[20] the authors compared color normalization techniques and color augmentation. They showed that both techniques improved generalization to unseen stain variations and that color augmentation led to better accuracy. However, a drop in performance with both techniques for the classification task on unseen center data was observed, which is not satisfactory for clinical use.

One of the most challenging aspects of computational pathology is the heavy size and weight of WSI (usually more than 10 000 x 10 000 pixels, and over 500 Mo, respectively). Traditional deep learning techniques using WSI as raw data are inapplicable, and when it comes to analyzing WSI, both storage and computation time are heavy challenges. The acceptance of diagnostic tools by pathologists requires the latter to be fit for usage and effective (i.e., fast and precise) in their daily routine practice. With these challenges in mind, we aimed to design a fast and simple processing pipeline that can be generalized to any center using a limited amount of training data, sticking to light model architecture for performance purposes.

Here we present a patch-based approach that detects and locates invasive carcinoma (IC) on WSI. The algorithm consists of a two-step pipeline. First, a filtering algorithm parses WSI regions that are relevant to IC detection, focusing the analysis on a small part of the slide and leading to faster processing. Filtered regions are partitioned into patches and fed into a convolutional neural network (CNN) based classifier that predicts whether those patches contain IC or not. This model was first trained with labeled data from a so-called "reference center", and performance was measured on a test dataset containing WSI from this same reference center. Scanners and staining methods for each center are summarized in Table 1. When testing this so-called "master model" (or "master classifier") on data from another center, termed "target center", performance appeared to be greatly decreased. To maintain a good performance level, we used the master classifier as a starting state for transfer learning with training data extracted from the target center. The obtained "calibrated model" (or "calibrated classifier") performed well on data from the target center.

Below, we first explain the WSI processing pipe by describing the filtering process and the architecture of the IC classifier. We then present the various datasets used in this study, both for training the classifier and testing the IC detection algorithm. We give insights on the labeling process and on the composition of the datasets. We then describe the training processes leading to the master classifier and the calibration step used to obtain the target classifier. The performance evaluation methodology is then described, allowing us to measure the system's efficiency both at patch level (i.e., how accurately IC is localized) and at slide level (i.e., whether a slide containing IC is predicted as such and vice-versa). Finally, the results obtained and future studies are discussed.

## METHODS

### Whole slide image pipeline processing

To locate IC on WSI we adopted a patch-based approach.[8,17,21] An input WSI was parsed into small images called patches which were filtered and later fed to a CNN-based classifier to determine whether they contained IC or not. From there a slide level score was determined to assess whether the slide contained IC or not. The full processing pipeline is illustrated in Figure 1.

In the parsing step, we chose to focus only on nuclear epithelial areas since these are the regions where IC diagnosis is performed. To do so, epithelial regions were first segmented automatically and were then parsed into patches at zoom level x20 and with size 256 x 256 pixels; non relevant patches: blur, no tissue, or those that did not contain nuclei were finally discarded. Doing so dramatically reduced the number of patches going through further processes; depending on the size of the epithelial regions filtering can discard up to 99% of the tissue (cf. Figure 1c). The filtering process is further described in Annex 1.

For each retained patch, a subsidiary patch with the same size and from the same center, but extracted at zoom level x5, was retrieved and fed into a two-class CNN-based classifier. A major difference with trad patch-based approaches,[17,21] was that we fed the network with x5 zoom patches to determine IC scores, these scores being attributed to the associated x20 zoom patches. The IC scores represent the probability that the patch contains IC (see section Evaluation method for more details). In other words, the contextual information contained in the x5 zoom patch determined the label of its central region (see Figure 1d, 1e). Keeping the x20 patch (and not the x5 one) as a base unit for the analysis allowed us to have a better resolution for cancer localization.

### Network architecture

The model architecture consisted of a Resnet50[22] feature extractor, the final fully connected layers being replaced by a random forest classifier. This hybrid architecture was found to improve results compared to using a simple ResNet50 architecture. The classifier assigned each patch with a probability of belonging to the IC class.

### Material and dataset

In this section the main libraries incorporated in our algorithm are mentioned and the datasets used for both the training and test are described.

### Material

The filtering step was performed on CPU units: Intel(R) Core(TM) i9 3.6 GHz.
Model training and IC inference were performed on GPU: NVIDIA GeForce RTX 2080 Ti, 11 Gb, frequency boost 11 GHz.

The training of the CNN was managed with tensorflow 2.5 and the training of the random forest classifier was managed with scikit-learn. To process WSI, we used the Openslide library.

**Datasets**

The datasets used in this study were built from WSI originating from two different medical centers referred to respectively as "reference center" and "target center". As described in this section, slides were first labeled, then patches were extracted and filtered to generate training and test datasets. The training datasets were used to train the networks and the test datasets were used to assess the network's performance. The complete extraction and training pipe is illustrated in Figure 2.

Both datasets were annotated according to the same process: using in-house software, 25 experienced pathologists were asked to draw regions of interest (ROI) on the WSI. Each ROI was checked and validated by an expert pathologist, in case of disagreement the WSI is sent back for revision until a consensus is found. ROI were drawn for regions presenting pathological patterns and for some healthy tissue as well. Each ROI was associated with a label corresponding to its diagnosis. Labels corresponding to invasive ductal carcinoma, invasive lobular carcinoma or mucinous carcinoma were grouped into the IC class and every other label corresponding to benign pathologies or healthy tissue was grouped into a Rest class.

Annotated slides were then processed and parsed into patches, each patch being assigned the label corresponding to the ROI it belonged to. Annotated ROIs may contain non-tissue pieces, blurry regions or stroma; to obtain quality data, those patches were filtered out according to the same criteria as those applied during the filtering step in the processing pipeline (see section "Whole slide processing pipeline"). The datasets are described in Table 2. For each center, several slides from the same patient were usually available. In order to split each dataset into a training and a test set, patient folders were divided so that ~80% of the slides lay in the training set. A given patient folder exclusively belonged to the test or the training set. This ensured that no data leakage occurred between the training and test sets.

As shown in Table 2, the reference dataset contained much more data than the target dataset; the former was used for training the master classifier whereas the target dataset was used for calibration.

**Training of the classification model**

In order to obtain an accurate classification model, relevant image features must be extracted from the images. This was achieved using a CNN that automatically learned optimal information for the task at hand, in this work the CNN used was a ResNet50. This network was a critical component of the proposed processing pipeline and was trained using a large amount of quality annotated data.

In [20], the authors show that data augmentation significantly improves models performance on histological data. Building on this knowledge, a similar set of data augmentation functions were used during the CNN training to reduce overfitting and improve results on unseen data, namely flips and rotations, additive Gaussian noise, hue, saturation, contrast, and brightness transformations. The various parameters controlling augmentation functions were set using a grid search scheme.

To train the ResNet50, binary cross entropy was optimized with an Adam optimizer[23] with an initial learning rate of 0.001. ResNet50 weights were initialized with weights resulting from pre-training on ImageNet and no layer was frozen. An early stopping condition was set. The features learnt from this feature extractor were used as inputs to train a random forest classifier. More specifically, the random forest was fed with feature maps output by the ResNet50 backbone from which the final fully connected layer was removed.

The network resulting from the training phase was the master model and could then be calibrated with data from another center in order to perform well on the latter.

**Calibration step to address a new center**

From one medical center or laboratory to another, slide preparation methods, staining processes, and slide scanner types can differ considerably. Because of these variations, the general aspect of the scanned slides is notably different (see Figure 3). Consequently, although the master model was shown to be robust to analyze data from the same center as the training dataset, its performance deteriorated when moving to a new acquisition center (see Results section). In order to address this problem, we set up a calibration step which made it possible to specifically adapt the master model to a new target center. The calibration step was based on transfer learning: weights from the master model were taken as the starting state for new training with labeled data from the target center. An almost 10-fold decrease in data required during the calibration step was observed when compared to the training on the reference center (see Table 3). The calibration training was performed with the same hyperparameters and training strategy as the master CNN training.

The target model resulting from this calibration step was especially trained to perform well on data from the target center. This model could be directly plugged into the pipeline of WSI processing described above, as illustrated in Figure 4.

**Evaluation method**

The performance of the master and calibrated models was measured on both reference and target centers. As described in "Material and dataset", patch test sets (i.e., reference test set and target test set) were built from slides from both centers to assess the generalizability of our algorithm.

The performance of each model was evaluated both at the patch and slide level. For patch level classification, a simple threshold of the IC score determined whether a given patch was classified as IC or Rest. This threshold was determined as the best precision/recall tradeoff by maximizing the F1-score on validation data. For slide level classification, we defined the slide score $S_{IC}$ as the sum of the scores of patches classified as IC within this slide normalized by the total number of patches detected on the WSI.

$$S_{IC} = \sum_{P>P_0} P/N,$$

where $P$ is the patch classification score, $P_0$ is the threshold from which a patch is positive to IC and $N$ is the total number of epithelial patches. If this score was above a fixed threshold (also determined using F1-score maximization) then the slide was considered to contain IC. The number of false negatives (FN), false positives (FP), true positives (TP), and true negatives (TN) were then computed both at the patch and slide level, and standard accuracy, precision, and recall metrics were measured. Accuracy was defined as (TP+TN)/(TP+FP+FN+TN) and represented the percentage of patches (resp. slides) correctly predicted; precision was defined as TP/(TP+FP) and represented the percentage of cancer-predicted patches (resp. slides) that were truly cancerous; recall was defined as TP/(TP+FN) and represented the percentage of patches (resp. slides) presenting IC regions that were correctly predicted as cancer.

**RESULTS**

The results are summarized in Table 4.

**Accuracy of the model**

At the patch level, the model had an accuracy of 92.1% and 96.3% on the reference and target test sets, respectively. At the slide level, accuracy was 97.6% for the reference test set and 92.0% for the target test set.

**Model recall**

At the patch level, the model had a recall of 95% and 87.8% on the reference and target test sets, respectively. At the slide level, recall was 90.9% for the reference test set and 100% for the target test set (Table 4).

**Precision of the model**

At the patch level, precision was 73.9% and 70.6% for the reference and target test sets, respectively. At the slide level, precision was 100% for test set 1 and 70.8% for test set 2 (Table 4).

**Computation times**

The average computation time of the complete pipeline for slides from our test set was 114.9 s with standard variation of 95 s. This could be broken down as follows: 108.3 s on average for the filtering process and 16.6 s on average for the IC inference step.

**DISCUSSION**

The algorithm developed and described in this paper aimed at detecting IC on WSI. It proved to be accurate and precise when tested at patch and slide level on the test datasets. However, to the best of our knowledge, there is no common standard either for metrics used or for datasets, and authors usually test their algorithms on their own data using different metrics. Some datasets do not even allow slide level evaluation and metrics are only computed at patch level. These reasons make any comparison complicated. For instance, Narayanan and colleagues[17] used a simple patch-based 5-layer CNN with color constancy pre-processing and achieved an AUC of 0.935 on a breast histology image dataset from Kaggle.[24] In another study, Cruz-Roa and colleagues introduced a 3-layer convnet classifier with a single convolutional layer.[25] A slide level DICE coefficient was computed to evaluate the model. These authors reported a DICE of $0.7494 \pm 0.2071$, positive predictive value (PPV) of $0.6464 \pm 0.2870$, and negative predictive value (NPV) of $0.9709 \pm 0.0350$ on a dataset consisting of TCGA breast images.[26] Celik and coworkers[8] were able to achieve 91.96% accuracy on the BreakHis dataset[27] at patch level using a pretrained ResNet-50 with no data-augmentation. Using a FusionNet encoder + softmax classifier, Brancati's group[28] reported 87.76% accuracy on Janowczyk's dataset.[29] Finally, Zeng and Zhang[7] used Google Cloud AutoML Vision on breast histopathology images from Mooney's dataset[24] and claimed 85.26% balanced accuracy at the patch level.

As described above, our technique led to a patch level accuracy of $> 90\%$ (see section Results) and a slide level accuracy of $> 92\%$. This is a state-of-the-art performance although, for previously mentioned reasons, a fair and closer comparison with other techniques is difficult.

A key point of the proposed algorithm pipeline is that it focuses the analysis solely on nuclear epithelial regions, leading to faster treatment, compatible with routine clinical use. Contrary to previous reports, where regions on which the analysis will be performed are determined through machine learning techniques, we propose simple, fast, and efficient filtering criteria based on standard image processing methods. No additional training is therefore needed. To the best of our knowledge, this is an original approach.

Another focus of our work was the ability of our model to generalize to different acquisition centers with various scanners and preparation processes that cause changes in slide appearance. The use of a two phase-training process provided a similar performance on the target test dataset compared to the reference test set while using much less training data. The target dataset was more than 10-times smaller than the reference dataset. It should also be noted that the diversity of pathologies found in the target dataset was smaller than that observed in the target dataset. Therefore, the reference dataset contained a lot of information that was not contained in the target dataset. It is important not to forget the information learnt during the first stage of training when transferring the knowledge to the target. As can be observed from the decreased performance on the reference center when using the calibrated CNN instead of the master CNN, it can be seen that some information is lost during the calibration phase (see Table 3). This problem will be addressed in a future study. Zaneta Swiderska-Chadaj et al.[19] proposed a technique using Cycle-GAN style normalization for a multicenter algorithm with good performance. They achieved similar accuracy on their development dataset and test datasets built using data originating from different centers. However, they used a training set containing data from several centers to train their classification model. With our proposed method, the network needs only one reference center dataset and calibration can then be performed using a small amount of data from a given target center. Furthermore, in order to reduce the computation time and the expectancy on hardware capabilities at inference, the number of models used and their size is minimal. It is therefore more adapted to use a single calibrated CNN instead of a Cycle-GAN normalization step followed by a CNN classification step.

## CONCLUSION

Our IC detection pipe is efficient and could be applied to WSI from different medical centers using a limited amount of additional data. This tool may help pathologists to make a more accurate and faster diagnosis and postoperative treatment planning. Such a support can be used for quickly screening slides in high-throughput laboratories, selecting slides needed for fast immunohistochemistry, making practice more consistent, or assisting in reporting. Future studies include a parametric study on the size of the target dataset, for various medical centers, to determine the minimum volume of data necessary to maintain performance when moving from the reference to a target center, improvement of the processing time, generalization to pathologies other than IC, and organs other than the breast. We will also investigate multi-domain adaptation as a means of obtaining proper generalization results with an even more limited amount of data when addressing a specific target center.

**Acknowledgements**

The authors thank all the pathologists involved in the process of labeling and all the stakeholders involved in the process of reviewing, editing and publishing this article.

**Author contributions**

R.P.[1], N.P.[1], M.S.[11] and S.S.[1] researched data for the article and drafted the manuscript. All authors contributed to reviewing and editing the manuscript.

**Competing Interests**

I have read the journal's policy and the authors of this manuscript have the following competing interests: S.S.[1] and M.S.[1-10] are co-founders and salaried employees of Primaa, R.P.[1] and N.P.[1] are salaried employees of Primaa.

**References**

1. Sung H, Ferlay J, Siegel RL, et al. Global Cancer Statistics 2020: GLOBOCAN estimates of incidence and mortality worldwide for 36 cancers in 185 countries. *CA Cancer J Clin* 2021; **71:**209–49.
2. Maung R. Pathologists' workload and patient safety. *Diagn Histopathol* 2016; **22:**283–7.
3. Metter DM, Colgan TJ, Leung ST, Timmons CF, Park JY. Trends in the US and Canadian pathologist workforces from 2007 to 2017. *JAMA Netw Open* 2019; **2:**e194337.
4. Blay JY, Boucher S, Le Vu B, et al. Delayed care for patients with newly diagnosed cancer due to COVID-19 and estimated impact on cancer mortality in France. *ESMO Open* 2021; **6:**100134.
5. Jahn SW, Plass M, Moinfar F. Digital pathology: advantages, limitations and emerging perspectives. *J Clin Med* 2020; **9:**3697.
6. Cui M, Zhang DY. Artificial intelligence and computational pathology. *Lab Investig J Tech Methods Pathol* 2021; **101:**412–22.
7. Zeng Y, Zhang J. A machine learning model for detecting invasive ductal carcinoma with Google Cloud AutoML Vision. *Comput Biol Med* 2020; **122:**103861.
8. Celik Y, Talo M, Yildirim O, Karabatak M, Acharya UR. Automated invasive ductal carcinoma detection based using deep transfer learning with whole-slide images. *Pattern Recognit Lett* 2020; **133:**232–9.
9. Han W, Johnson C, Warner A, et al. Automatic cancer detection on digital histopathology images of mid-gland radical prostatectomy specimens. *J Med Imaging (Bellingham)* 2020; **7:**047501.
10. Bulten W, Pinckaers H, van Boven H, et al. Automated deep-learning system for Gleason grading of prostate cancer using biopsies: a diagnostic study. *Lancet Oncol* 2020; **21:**233–41.
11. Hekler A, Utikal JS, Enk AH, et al. Deep learning outperformed 11 pathologists in the classification of histopathological melanoma images. *Eur J Cancer* 2019; **118:**91–6.
12. Coudray N, Ocampo PS, Sakellaropoulos T, et al. Classification and mutation prediction from non-small cell lung cancer histopathology images using deep learning. *Nat Med* 2018; **24:**1559–67.
13. Wang KS, Yu G, Xu C, et al. Accurate diagnosis of colorectal cancer based on histopathology images using artificial intelligence. *BMC Med* 2021; **19:**76.

14. Jannesari M, Habibzadeh M, Aboulkheyr H, et al. Breast cancer histopathological image classification: a deep learning approach. In: 2018 IEEE International Conference on Bioinformatics and Biomedicine (BIBM). 2018. p. 2405–12.
15. Ehteshami Bejnordi B, Mullooly M, Pfeiffer RM, et al. Using deep convolutional neural networks to identify and classify tumor-associated stroma in diagnostic breast biopsies. *Mod Pathol* 2018; **31:**1502–12.
16. Dong F, Irshad H, Oh E-Y, et al. Computational pathology to discriminate benign from malignant intraductal proliferations of the breast. *PloS One* 2014; **9:**e114885.
17. Narayanan BN, Krishnaraja V, Ali R. Convolutional neural network for classification of histopathology images for breast cancer detection. In: 2019 IEEE National Aerospace and Electronics Conference (NAECON). 2019. p. 291–5.
18. Pantanowitz L, Quiroga-Garza GM, Bien L, et al. An artificial intelligence algorithm for prostate cancer diagnosis in whole slide images of core needle biopsies: a blinded clinical validation and deployment study. *Lancet Digit Health* 2020; **2:**e407–16.
19. Swiderska Z, Bel T, Blanchet L, et al. Impact of rescanning and normalization on convolutional neural network performance in multi-center, whole-slide classification of prostate cancer. *Sci Rep* 2020; **10:**14398.
20. Tellez D, Litjens G, Bandi P, et al. Quantifying the effects of data augmentation and stain color normalization in convolutional neural networks for computational pathology. *Med Image Anal* 2019; **58:**101544.
21. Cruz-Roa A, Basavanhally A, González F, et al. Automatic detection of invasive ductal carcinoma in whole slide images with convolutional neural networks. 2014; **9041:**904103.
22. Xie S, Girshick R, Dollár P, Tu Z, He K. Aggregated residual transformations for deep neural networks. ArXiv161105431 Cs [Internet]. 2017. [Accessed 7 Sept 2021]; Available at: http://arxiv.org/abs/1611.05431
23. Kingma DP, Ba J. Adam: A method for stochastic optimization. ArXiv14126980 Cs [Internet]. 2017. [Accessed 7 Sept 2021]; Available at: http://arxiv.org/abs/1412.6980
24. Breast Histopathology Images [Internet]. [Accessed 7 Sept 2021]. Available at: https://kaggle.com/paultimothymooney/breast-histopathology-images
25. Cruz-Roa A, Gilmore H, Basavanhally A, et al. Accurate and reproducible invasive breast cancer detection in whole-slide images: A deep learning approach for quantifying tumor extent. *Sci Rep* 2017; **7:**46450.
26. The Cancer Genome Atlas Program - National Cancer Institute [Internet]. 2018. [Accessed 7 Sept 2021]. Available at: https://www.cancer.gov/about-nci/organization/ccg/research/structural-genomics/tcga
27. Breast Cancer Histopathological Database (BreakHis) - Laboratório Visão Robótica e Imagem [Internet]. 2017. [Accessed 7 Sept 2021]. Available at:

https://web.inf.ufpr.br/vri/databases/breast-cancer-histopathological-database-breakhis/

28. Brancati N, De Pietro G, Frucci M, Riccio D. A deep learning approach for breast invasive ductal carcinoma detection and lymphoma multi-classification in histological images. *IEEE Access* 2019; **7:**44709–20.

29. CHOOSEHAPPY. Use Case 6: Invasive Ductal Carcinoma (IDC) Segmentation [Internet]. Andrew Janowczyk. 2015. [Accessed 7 Sept 2021]. Available at: http://www.andrewjanowczyk.com/use-case-6-invasive-ductal-carcinoma-idc-segmentation/

30. Nobuyuki Otsu. "A Threshold Selection Method from Gray-Level Histograms". In: IEEE Transactions on Systems, Man, and Cybernetics 9.1 (Jan. 1979), pp. 62–66. doi: 10.1109/tsmc.1979.4310076. url: https://doi.org/10.1109/tsmc.1979.4310076.

31. Sandler, Mark & Howard, Andrew & Zhu, Menglong & Zhmoginov, Andrey & Chen, Liang-Chieh. (2018). MobileNetV2: Inverted Residuals and Linear Bottlenecks. 4510-4520. 10.1109/CVPR.2018.00474.

**Annex 1. Filtering process**

As described in section Methods, a filtering process is applied to the WSI with a two-fold objective: reduce the analysis to epithelial regions and discard patches that have no interest for further analysis (artifacts, no tissue, no nuclei). This allows dramatically reducing the amount of patches fed to the invasive cancer classifier.

The filtering process is made of the following steps:

- epithelial regions segmentation
- discard patches that have no interest for further analysis
    - patches do not contain nuclei
    - blurry patches
    - patches with little tissue inside

The epithelial segmentation process is described in Figure 5, the tissue is first segmented at zoom x1 through a simple two class otsu[30] thresholding. The resulting mask is then parsed into 256*256px tissue patches at zoom x2.5. Epithelial regions appear as dark heterogeneous regions (see Figure 5). Tissue patches undergo a gaussian smoothing so as to make epithelial regions more uniform and a final two class otsu thresholding is applied to discriminate epithelium from stroma.

Epithelial tissue is then parsed at zoom x20 into 256*256px patches which go through a final discard process, each filter is described in Table 5.

**Table 1: Scanners and staining methods**

|  | Reference center | Target center |
|---|---|---|
| Scanner | Hamamatsu (S120) | 3DHistech (P1000) |
| resolution (μm/pixel) | 0.44 | 0.24 |
| Staining method | HES | HES |

**Table 2: Data set constituents**

| | Reference dataset | | | | | | Target dataset | | | | | |
|---|---|---|---|---|---|---|---|---|---|---|---|---|
| | Training set | | | Test set | | | Training set | | | Test set | | |
| Class | No. of patient folders | No. of slides | No. of patches | No. of patient folders | No. of slides | No. of patches | No. of patient folders | No. of slides | No. of patches | No. of patient folders | No. of slides | No. of patches |
| IC | 50 | 175 | 435913 | 7 | 17 | 49089 | 7 | 27 | 56586 | 4 | 22 | 15196 |
| Rest | 98 | 777 | 855366 | 18 | 71 | 137591 | 11 | 45 | 19481 | 10 | 61 | 182731 |
| Total | 148 | 952 | 1291279 | 25 | 88 | 186680 | 18 | 72 | 76067 | 14 | 83 | 197927 |

This table summarizes, for each class, the number of patient folders, slides, and patches that are contained in both datasets (training and test subsets). Note that a slide is referred to as an "IC slide" if it contains at least one region of interest with a label being one of the following: invasive ductal carcinoma, invasive lobular carcinoma or mucinous carcinoma. Dataset 1 (resp. 2) contains 135 (resp. 48) slides with ductal IC, 52 (resp. 0) slides with lobular IC, and five (resp. 1) slides with mucinous carcinoma. Note that slides come from both biopsies and mastectomies, in comparable proportions. The proportion of cases between various labels was not chosen, it is representative of the routine practice of the medical centers that provided the data.

**Table 3: Comparison of master model and calibrated model performance**

|  | **Master model** | **Calibrated model** |
|---|---|---|
| **Reference center test set** | **Accuracy: 0.92**<br><br>**Precision: 0.94**<br><br>**Recall: 0.74** | Accuracy: 0.75<br><br>Precision: 0.44<br><br>Recall: 0.66 |
| **Target center test set** | Accuracy: 0.51<br><br>Precision: 0.95<br><br>Recall: 0.07 | **Accuracy: 0.96**<br><br>**Precision: 0.87**<br><br>**Recall: 0.70** |

This table illustrates the results of both the master IC classifier model and the target IC classifier model, on both the reference and target test sets. The master model performs well on the reference test set but its performance is greatly decreased when translating to the target center. Differences in slides preparation and acquisition scanners make it challenging to generalize to a new center. The calibrated model is obtained through a transfer learning approach by taking the master model as the starting state when learning on a training dataset from the target center. The performance of the target model obtained on the target test set is highly improved and is comparable to that obtained by the master model when tested on the reference test. Note that the target model has its metrics decreased on the reference test set.

**Table 4. Results of the model on test sets 1 and 2**

|  |  | **Reference dataset Test set** | **Target dataset Test set** |
|---|---|---|---|
| **Patch level metrics** | Accuracy | 0.92 | 0.96 |
|  | Recall | 0.94 | 0.87 |
|  | Precision | 0.74 | 0.70 |
| **Slide level metrics** | Accuracy | 0.98 | 0.92 |
|  | Recall | 0.91 | 1.0 |
|  | Precision | 1.0 | 0.71 |

The table details the classification metrics at patch and slide levels for the reference and target datasets. This illustrates the calibrated system's performance.

**Table 5: filtering functions**

| filter | algorithm | details |
| --- | --- | --- |
| discard patches that do not contain nuclei | MobileNetV2 classifier[31] | The network was trained to distinguish patches that contain nuclei from patches that do not contain nuclei |
| blur | a laplacian filter is applied to the image. If the variance of the resulting mask is under a given threshold, then the image is filtered out | Patches that are too blurry are filtered out |
| not enough tissue | - gets the most frequent pixel value **val** in the image converted to greyscale<br>- checks the proportion **prop** of pixels with intensity **i** such as \| **val - i**\| < **thres_1**<br>- discards the patch if **prop** > **thres_2** | Patches that do not contain enough tissue are filtered out |

The table summarizes the various filters applied in the filtration process. Patches that do not contain nuclei, that are blurry or that contain too little tissue inside are filtered out.

**Figure 1. Whole slide image (WSI) processing pipeline for invasive carcinoma (IC) detection.**

This figure illustrates the processing pipeline for invasive carcinoma (IC) detection on breast WSI (a). A zoomed-in view of the tissue is shown in (b). Epithelial nuclear regions are first detected and parsed (c) into square patches at zoom x20 (e). For each x20 patch, an auxiliary patch with the same center and size is extracted at zoom x5 (d) and fed to a CNN based classifier to predict an IC score. This score is attributed to the x20 patch. The set of IC scores for x20 patches belonging to nuclear epithelial regions enables the location of cancer on a slide (f). Note that a patch is considered to fall into the IC class when its score is above a threshold determined through the ROC method. NB: the color code in (f) ranges from blue (very low IC score - low probability of IC) to red (very high IC score - high probability of IC). A slide IC score is then computed by taking the weighted average of the IC patches score.

**Figure 2. Two-stage dataset generation and training process.**

This figure illustrates our two stage dataset generation training process. Whole slide images (WSI) from the reference center are parsed and broken down into patches. These are filtered so as to keep only those belonging to nuclear epithelial regions. The patches obtained are split between the training and test sets (at a ratio of 0.8, resp 0.2). Note that a given patient's slides are either fully included in the test or training set so as to avoid data leakage. The reference training set is used to train the master IC classifier. The performance of the master classifier is evaluated on the reference test set.
Regarding the calibration process, the master model is used as an initial state for transfer learning training on the target center training set. The resulting calibrated model is evaluated on the target test set.

**Figure 3. Variations in slide appearance.**

This figure shows two slides from the reference and target centers. Due to variations in the slide preparation process and acquisition scanners from one center to another, the slide aspect (i.e tissue staining and texture) can differ. This results in problems of generalization when a model that has learnt on a given center is used for inference on WSI from another center.

**Figure 4. Development of the deep learning model using two-phase training.**

This figure illustrates how the IC classifier models are plugged into the WSI processing pipeline. Any classifier can be plugged into the pipeline. When implemented in a target center, the plugged model is calibrated on the target center training set, using the master CNN model as a starting state for training.

**Figure 5. Epithelial regions segmentation**

This figure illustrates the WSI epithelium segmentation process. The tissue is first segmented at zoom 1 through a simple two class otsu thresholding. The resulting mask is then parsed into 256*256px tissue patches at zoom 2.5. Epithelial regions appear as dark heterogeneous regions. Tissue patches undergo a gaussian smoothing so as to homogeneous epithelial regions and a final two class otsu thresholding is applied to discriminate epithelium from stroma.